\documentclass[prl,twocolumn,aps]{revtex4}
\usepackage{graphicx}   
\usepackage{epstopdf}

\begin{document}

\title{A repulsive reference potential reproducing the dynamics of a liquid with attractions}

\author{Ulf R. Pedersen}\email{urp@berkeley.edu}
\affiliation{Department of Chemistry, University of California, Berkeley, California 94720-1460, USA}
\affiliation{DNRF centre ``Glass and Time,'' IMFUFA, Department of Sciences, Roskilde University, Postbox 260, DK-4000 Roskilde, Denmark}
\author{Thomas B. Schr{\o}der}\email{tbs@ruc.dk}
\affiliation{DNRF centre ``Glass and Time,'' IMFUFA, Department of Sciences, Roskilde University, Postbox 260, DK-4000 Roskilde, Denmark}
\author{Jeppe C. Dyre}\email{dyre@ruc.dk}
\affiliation{DNRF centre ``Glass and Time,'' IMFUFA, Department of Sciences, Roskilde University, Postbox 260, DK-4000 Roskilde, Denmark}

\date{\today}

\begin{abstract}
A well-known result of liquid state theory is that the structure of dense fluids is mainly determined by repulsive forces. The WCA potential, which cuts intermolecular potentials at their minima, is therefore often used as a reference. However, this gives quite wrong results for the viscous dynamics of the Kob-Andersen binary Lennard-Jones liquid [Berthier and Tarjus, Phys. Rev. Lett. {\bf 103}, 170601 (2009)]. This paper shows that  repulsive inverse-power law potentials provide a useful reference for this liquid by reproducing its structure, dynamics, and isochoric heat capacity.
\end{abstract}

\maketitle

Condensed matter is held together by attractive intermolecular forces, but is generally very resistant against compression. These two simple observations provide important microscopic information, and long ago it was shown that many properties of liquids and solids derive directly from the fact that intermolecular forces consist of short-ranged, harsh repulsions and long-ranged, much weaker attractions \cite{vdw1873,wid67,cha83}. This is the background of the noted Weeks-Chandler-Andersen (WCA) papers \cite{cha83,wee71} in which a repulsive reference potential is arrived at by cutting off the potentials at their minima; similar notable works emphasizing the dominance of the repulsive interactions were published by Widom \cite{wid67}, Barker and Henderson \cite{bar67}, and Gubbins {\it et al.} \cite{gub71}. During the 1970's it was demonstrated that the WCA reference represents well the liquid structure as monitored, e.g., by the radial distribution function \cite{cha83,wee71}. Since then thermodynamic perturbation theories \cite{books} start from the insight that intermolecular attractions give rise merely to an almost constant mean-field background potential.

There are several examples where the WCA reference reproduces the liquid dynamics, but exceptions to this have also been reported \cite{wcadyn}. In a striking case, Berthier and Tarjus recently showed by simulation that the dynamics of the Kob-Andersen binary Lennard-Jones (KABLJ) viscous liquid is not at all properly reproduced by the WCA reference \cite{ber09}. Berthier and Tarjus concluded that, while the attractive forces have little effect on the liquid's structure, they affect the dynamics in a highly nontrivial and nonperturbative way. In view of these finding an obvious question is whether other repulsive potentials can better reproduce the KABLJ liquid's dynamics. We focus here on inverse power-law (IPL) potentials, which are well known to have a number of simple properties \cite{ipl}. We simulated three Kob-Andersen type binary mixtures: the original LJ version (KABLJ) \cite{kob94}, the WCA version (KABWCA), and an inverse power-law version (KABIP). The KABLJ liquid is an 80:20 mixture of two LJ particles, A and B \cite{kob94}. This liquid has become a standard system for studying viscous liquid dynamics, and like most binary mixtures with size ratio below 0.9 it doesn't easily crystallize \cite{bernu,tox09}.  If $r_{ij}$ is the distance between particles $i$ and $j$, the KABLJ potential energy is given by

\begin{equation}\label{kablj_pot}
U\,=\,
4\sum_{i>j}  \varepsilon_{ij}\left[\left(\frac{\sigma_{ij}}{r_{ij}}\right)^{12}-\left(\frac{\sigma_{ij}}{r_{ij}}\right)^{6}\right]\,.
\end{equation}
The indices $i$ and $j$ vary over all particles, and the energy and length parameters are given as one of three combinations: $\varepsilon_{AB}/\varepsilon_{AA}=1.5$ and $\varepsilon_{BB}/\varepsilon_{AA}=0.5$, $\sigma_{AB}/\sigma_{AA}=0.8$ and $\sigma_{BB}/\sigma_{AA}=0.88$ \cite{kob94}.

\begin{figure}
\begin{center}
\includegraphics[width=0.85\columnwidth]{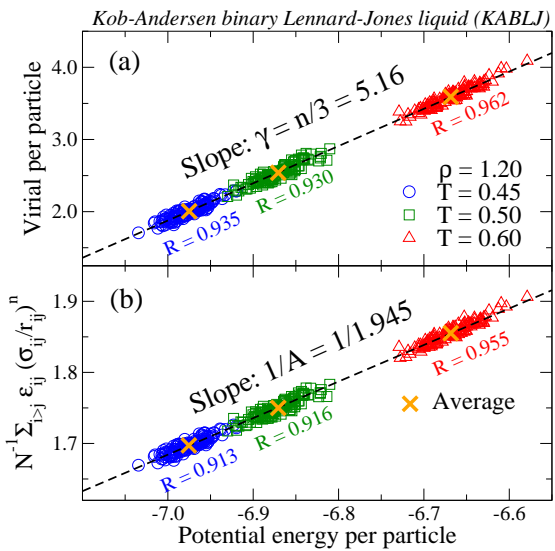} \newline
\end{center}
\caption{\label{figUncover} (color online) (a) The IPL exponent is estimated to be $n=3\times 5.16 = 15.48$ by evaluating the slope of a scatter plot of the virial and potential energy fluctuations of the KABLJ liquid at three state points with same density (open symbols, NVT simulations of 1000 particles). Average values of the state points are indicated by yellow $\times$'s. (b) The IPL prefactor is estimated to be $A=1.945$ from the slope in a scatter plot of  $\sum_{i>j}\varepsilon_{ij}(r_{ij}/\sigma_{ij})^{-n}$ and potential energy fluctuations. The number $R$ gives the virial / potential energy correlation coefficient in the NVT ensemble.}
\end{figure}

We wish to compare the dynamics of the KABLJ liquid to that of an IPL liquid. A simple IPL ansatz that respects the difference between A and B particles is

\begin{equation}
U_\textrm{IPL}\,=\,
A\sum_{i>j}\varepsilon_{ij}\left(\frac{\sigma_{ij}}{r_{ij}}\right)^{n}\,.
\end{equation}
This ansatz leaves just two parameters to be determined, $A$ and $n$. For an IPL liquid with $r_{ij}^{-n}$ interactions the virial obeys $W=(n/3)U$ for all microscopic states \cite{ipl}. Figure \ref{figUncover}(a) shows constant-volume equilibrium fluctuations of virial versus potential energy of the KABLJ liquid at three state points with same density in the supercooled regime (the three colored ovals) \cite{rumd}. The slope $\gamma$ of the best fit line is $5.16$. Thus we choose $n=3\times 5.16 = 15.48$ as the IPL exponent \cite{gamma_note}. Once $n$ is fixed, $A$ is determined by plotting the fluctuations of the KABLJ potential energy versus $\sum_{i>j}\varepsilon_{ij}({{\sigma_{ij}}/r_{ij}})^{n}$, referring to configurations drawn from the KABLJ simulations (Fig. \ref{figUncover}(b)). This plot yields $A=1.945$ as the value ensuring overall agreement between the magnitudes of the KABLJ and the KABIP potential energy fluctuations \cite{note}.

\begin{figure}
\begin{center}
\includegraphics[width=0.85\columnwidth]{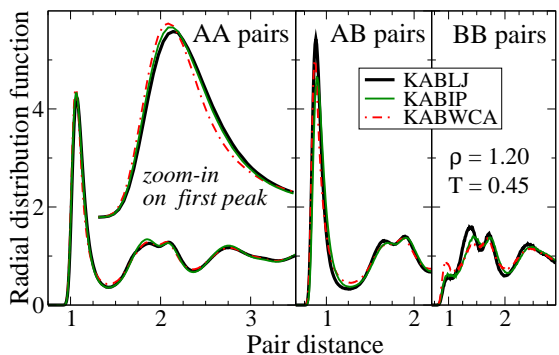}
\end{center}
\caption{\label{FigGr} (color online) Radial distribution functions of the AA, AB, and BB particle pairs of the KABLJ, KABIP, and KABWCA liquids at one state point.}
\end{figure}

Before comparing the dynamics of the KABLJ and KABIP liquids we briefly consider their structure by evaluating the radial distribution functions (RDF) in Fig. \ref{FigGr}. This figure also includes the RDF of the KABWCA liquid. The IPL is slightly better than the WCA, but overall the three potentials give quite similar RDFs. The largest deviations are seen for the small particle (BB) distribution; note that all three potentials capture its characteristic small first peak (with peak value below unity) stemming from the strong AB affinity.

\begin{figure}
\begin{center}
\includegraphics[width=0.85\columnwidth]{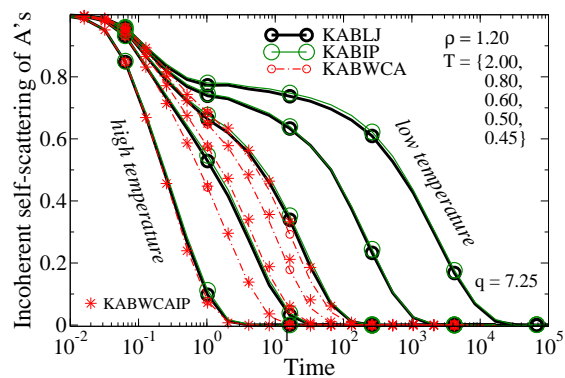} \newline
\end{center}
\caption{\label{FigFs} (color online) Incoherent intermediate self-scattering function for the A particles of the KABLJ, KABIP, and KABWCA liquids along the $\rho=1.20$ isochore for five temperatures. At the highest temperature the three models have the same dynamics, but at lower temperatures only the KABLJ and KABIP liquid have the same dynamics; the KABWCA liquid is considerably faster. Asterices (KABWCAIP) denote the IPL approximation to the KABWCA liquid (at each state point the parameters $A$ and $n$ are uniquely determined from the equilibrium virial and potential energy fluctuations, resulting in at $\rho=1.20$ for $T=\{0.45,0.5,0.6,0.8,2.0\}$: $n=\{20.67, 20.28, 19.59, 18.69, 16.38\}$ and $A=\{1.176,1.205,1.231,1.267,1.471\}$).}
\end{figure}

\begin{figure}
\begin{center}
\includegraphics[width=0.85\columnwidth]{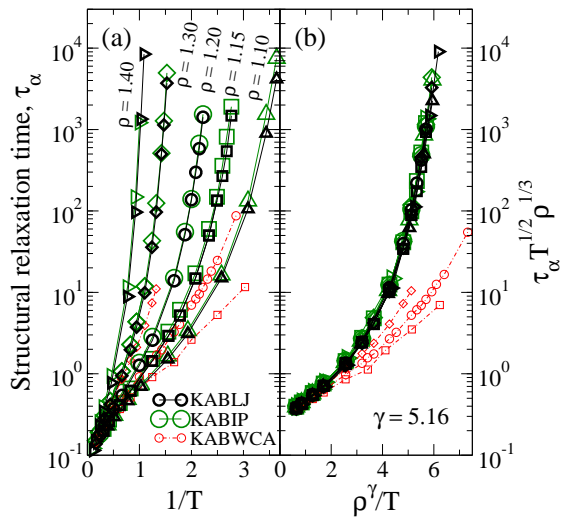}
\end{center}
\caption{\label{FigTauAlpha} (color online) (a) Structural relaxation time $\tau_\alpha$ as a function of inverse temperature defined as the time when the AA incoherent intermediate scattering function is $1/e$ at wave vector $q=7.25(\rho/1.2)^{1/3}$. Lines between isochoric points are shown for clarity. (b) The reduced structural relaxation time of the KABLJ and KABIP liquids is a function of the variable $\rho^\gamma/T$, confirming density scaling \cite{ds,cos}. In both (a) and (b) the red symbols give the KABWCA results at densities 1.15, 1.20, and 1.30.}
\end{figure}

Figure \ref{FigFs} compares the KABLJ, KABIP, and KABWCA dynamics via the A particle incoherent intermediate self-scattering function evaluated at the wave vector of the first peak of the AA structure factor, $q=7.25$ (reduced units referring to the A particle parameters are used throughout). The simulations were performed over a range of temperatures on the $\rho=1.20$ isochore, giving more than four decades of variation in the structural relaxation time. The agreement between the KABLJ and KABIP dynamics is good. The intermediate scattering function of the KABWCA liquid likewise agrees with that of the KABLJ liquid at the highest temperature, but the KABWCA dynamics is too fast at low temperatures \cite{ber09}. Comparing Figs. \ref{FigGr} and \ref{FigFs} one concludes that the RDFs do not determine the dynamics. That pair correlation functions do not determine the dynamics may be relevant, for instance, for illuminating the applicability of mode-coupling theories \cite{ber10}.  As an aside, Fig. \ref{FigFs} includes also a system (KABWCAIP) approximating the KABWCA system by IPL potentials. As for the KABLJ liquid the two parameters $A$ and $n$ were determined (uniquely) from the equilibrium fluctuations (but in this case, interestingly, the parameters vary with state point whereas they are virtually constant for the KABLJ liquid).

The structural relaxation time $\tau_\alpha$ is defined here as the time where the AA incoherent self-scattering function is $1/e$. This quantity's temperature dependence is shown in Fig. \ref{FigTauAlpha}(a). We here  varied the density between 1.10 and 1.40, corresponding to pressures between -1 and 11 kbar in the supercooled viscous regime (argon units). Again there is good agreement between the KABLJ and KABIP systems (Fig. \ref{FigTauAlpha}(a)). Figure \ref{FigTauAlpha}(b) shows that the KABLJ liquid obeys density scaling \cite{ds,cos}, i.e.,  that the reduced structural relaxation time is a function of $\rho^\gamma/T$. In particular, this implies that the IPL reference potential also conforms well to the fragility scaling of Berthier and Tarjus (Fig. 2 in Ref. \onlinecite{ber09}).

The simulation results for $\tau_\alpha(T)$ may be fitted by several well-known analytical functions for the temperature dependence along an isochore, both by functions with a dynamic divergence (e.g., the Vogel-Fulcher equation) and by functions without a dynamic divergence (e.g., quadratic scaling \cite{elm09} or other recent suggestions \cite{hec08,mau09}). For clarity of presentation we did not plot these functions. In all cases density scaling dictates the density dependence of the scaling parameters with the exponent $\gamma$ determined from simulations at one density (Fig. \ref{figUncover}).

\begin{figure}
\begin{center}
\includegraphics[width=0.85\columnwidth]{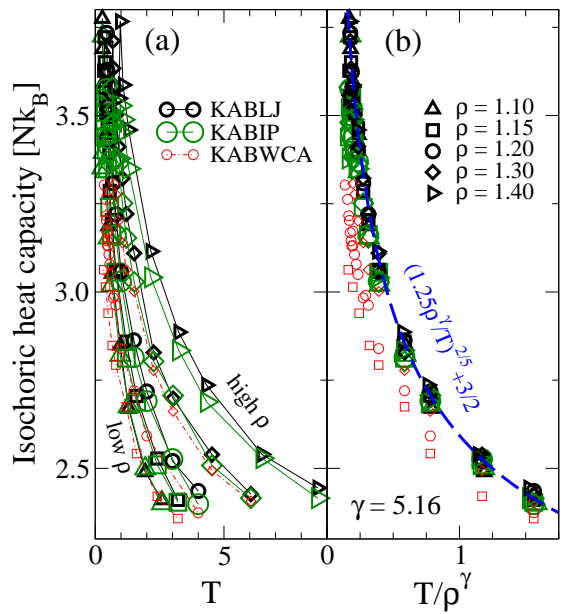}
\end{center}
\caption{\label{cv} (color online) (a) Isochoric specific heat per particle, $c_V$, in units of $k_B$. Lines between isochoric points are shown for clarity. The KABIP $c_V$ is closer to the KABLJ than is the KABWCA. (b) The same data scaled according to the density scaling recipe in conjunction with the Rosenfeld-Tarazona $c_V$ scaling \cite{ros98}, which is indicated analytically by the blue dashed line (the number 1.25 being a fitting parameter).}
\end{figure}

Finally, we show in Fig. \ref{cv}(a) the isochoric specific heat per particle, $c_V$, as a function of temperature for the three potentials at different densities. Fig. \ref{cv}(b) shows that the KABLJ and the KABIP liquids both conform to density scaling as well as to Rosenfeld-Tarazona's $c_V$ scaling ($c_V-3/2\propto  T^{-2/5}$) \cite{ros98}. The KABLJ and KABIP liquids agree, while the KABWCA liquid gives somewhat smaller $c_V$'s. Thus not only for the dynamics, but also for a static thermodynamic quantity like $c_V$ does an IPL system provide a superior reference system.

Why does an IPL potential so well reproduce the KABLJ liquid's dynamics? It was recently argued that van der Waals-type liquids have a ``hidden'' (approximate) scale invariance, which is reflected in the fact that these liquids to a good approximation inherit a number of properties of repulsive inverse power law (IPL) potentials \cite{scl}. As mentioned, an IPL potential exhibits 100\% correlation between virial $W$ and potential energy $U$ for all microscopic configurations [$W=(n/3)U$]. Hidden scale invariance is manifested in the appearance of strong correlations of the thermal equilibrium $WU$ fluctuations in the $NVT$ ensemble (Fig. \ref{figUncover}); it comes about as follows \cite{scl}. Highly asymmetric potentials like the LJ pair potential are well approximated around the minimum by an ``extended inverse power law'' (eIPL) of the form $v(r)=Ar^{-n}+B+Cr$. When summing over all nearest-neighbor distances, the linear term almost doesn't fluctuate if the total volume is fixed. Consequently, several properties of strongly correlating liquids are well reproduced by  IPL potentials. It should be noted, though, that because the linear term of the eIPL potential gives a volume-dependent contribution to the free energy, the IPL free energy, pressure, compressibility, and equation of state, represent poorly the liquid \cite{scl}.

Strongly correlating liquids include the van der Waals and metallic liquids, but neither covalently and hydrogen-bonded liquids, nor strongly ionic liquids \cite{scl} -- this is because competing interactions generally spoil strong $WU$ correlations. To a good approximation strongly correlating liquids obey  density scaling \cite{ds} according to which the relaxation time at varying temperatures and pressures is a function of $\rho^\gamma/T$. In simulations of strongly correlating liquids $\gamma$ is determined from the approximate identity $\Delta W(t)\cong\gamma\Delta U(t)$ (Fig. \ref{figUncover}) \cite{scl,cos}. Generally, strongly correlating liquids are characterized by the existence of ``isomorphic'' curves in their phase diagram along which several quantities are almost invariant, e.g., the excess entropy and the relaxation time in reduced units \cite{iso}. Whenever $\gamma$ is fairly constant throughout large parts of the phase diagram -- as appears to be the case for most strongly correlating liquids \cite{scl} -- an isomorph is to a good approximation given by $\rho^\gamma/T={\rm Const}$. This explains density scaling for the class of strongly correlating liquids \cite{iso}. 

In conclusion, structure, dynamics, and some thermodynamic properties of the KABLJ liquid are well represented by a repulsive IPL reference. Our findings suggest that the WCA reference fails to reproduce the viscous dynamics because this reference does not incorporate the hidden scale invariance of the KABLJ liquid. 

------------

The authors are indebted to David Chandler, Peter Harrowell, Ken Schweizer, and S{\o}ren Toxv{\ae}rd for fruitful discussions during the preparation of this paper. 
URP  is supported by the Danish Council for Independent Research in Natural Sciences. 
The centre for viscous liquid dynamics ``Glass and Time'' is sponsored by the Danish National Research Foundation (DNRF).


\begin{thebibliography}{99}
\bibitem{note}
Alternatively, the numbers $A$ and $n$ may be determined from temperature derivatives of the thermodynamic averages: Thus the yellow $\times$'s on Figs. \ref{figUncover}(a) and \ref{figUncover}(b) indicate that $A$ and $\gamma=n/3$ are determined from the averages. First $n$ is fixed to get the correct $\beta_V^{\rm ex}/C_V^{\rm ex}$ [where $\beta_V^{\rm ex}\equiv(\partial W/\partial T)_V$]. Then $A$ is fixed by requiring that the correct $C_V^{\rm ex}$ is reproduced when evaluated for the IPL potential for the configurations visited when using the full potential.
\end{thebibliography}
\end{document}